\documentclass[journal]{IEEEtran}
\usepackage{cite,color}

\usepackage[dvips]{graphicx}

\usepackage{amsmath}

\begin{document}
%
\title{Iterative image reconstruction for CT with unmatched projection matrices using\\
the generalized minimal residual algorithm}

\author{Emil~Y.~Sidky, Per~Christian~Hansen, Jakob~S.~J{\o}rgensen, and Xiaochuan~Pan
\thanks{E. Y. Sidky and X. Pan are with the Department
of Radiology, University of Chicago, Chicago,
IL, 60637 USA. emails: sidky@uchicago.edu, xpan@uchicago.edu}
\thanks{P. C. Hansen and J. S. J{\o}rgensen are with the Department of Applied Mathematics
and Computer Science, Technical University of Denmark, Kgs. Lyngby, DK-2800, Denmark.
emails: pcha@dtu.dk, jakj@dtu.dk}}

\maketitle
\thispagestyle{empty}

\begin{abstract}
The generalized minimal residual (GMRES) algorithm
is applied to image reconstruction using linear computed tomography (CT) models.
The GMRES algorithm iteratively
solves square, non-symmetric linear systems and it
has practical application to CT when using unmatched back-projector/projector
pairs and when applying preconditioning.
The GMRES algorithm is demonstrated on a 3D CT image reconstruction problem
where it is seen that use of unmatched projection matrices
does not prevent convergence, while
using an unmatched pair 
in the related conjugate gradients for least-squares (CGLS) algorithm
leads to divergent iteration.
Implementation of preconditioning using GMRES is also demonstrated.
\end{abstract}


%
\IEEEpeerreviewmaketitle

\section{Introduction}
%
%
%
%
\IEEEPARstart{L}{inear} models for computed tomography (CT)
play 
an important role for iterative image reconstruction.
The most common
approach to 
CT processing involves taking the negative logarithm
of the projection data, so that the line integration model
leads to a linear relation between 
the image and processed data. Accordingly,
the CT image reconstruction problem can be written as a large
linear system
\begin{equation}
\label{axb}
Ax=b,
\end{equation}
where $b$, a vector of length $m$, represents the processed projection data;
$x$, a vector of length $n$, contains the image pixel values; and the $m\times n$
system matrix $A$ contain the weights that model line-integration.
Linear tomographic models can include quadratic regularization,
cf.\ \cite{Fessler94} and \cite[Chapter~12]{SIAMbook},
or more sophisticated modeling
such as noise correlation and blur due to accurate detector physics \cite{Tilley17}.
Even when non-linear models for CT \cite{Elbakri02} are considered for iterative
image reconstruction, there is usually a large linear system that is involved in the
algorithm. 
Novel techniques for solving large linear systems may thus be of practical
use for iterative image reconstruction in CT.

The most common iterative algorithm for solving linear CT models, excluding
row-action, sequential, or SIRT-type data processing methods, is the conjugate
gradients (CG) algorithm \cite[Chapter~11]{SIAMbook}.
For least-squares problems with non-symmetric
system matrices, in particular, there
is the conjugate-gradients least-squares (CGLS) algorithm that solves the
optimization problem
\begin{equation}
\label{cglsopt}
\min_x \frac{1}{2} \| Ax -b \|^2_2.
\end{equation}
The minimizer of this optimization problem can also be found from solving the
normal equations directly
\begin{equation}
\label{cgnormal}
A^\top A x = A^\top b,
\end{equation}
which are derived from \eqref{cglsopt} by taking the gradient
of the objective function and setting it to zero.
In applying CGLS, the implementation for back-projection $B$
must be the matrix transpose $A^\top$.
If $B \neq A^\top$ then the resulting method is not well defined, there is no
convergence theory, and if it does converge it
does not solve Eqs.\ (\ref{cglsopt}) and (\ref{cgnormal}).
Nevertheless there are practical motivations for considering back-projection
implementations $B$  different than $A^\top$.
These motivations are outlined in Ref. \cite{ZengGullberg} in connection
with SIRT-type iterative solvers,
where the authors explain that $B$ can be a preconditioner, $B$ may be an efficient
but approximate implementation of $A^\top$, or $A$ may involve complex physics
modeling that may make computer implementation of $A^\top$ prohibitively expensive.
As shown in \cite{dhhr2019fixing} we can guarantee convergence of
SIRT-type methods with $B \neq A^\top$
(but not CGLS) by shifting the complex eigenvalue spectrum of $BA$ so that eigenvalues
with negative real part
are eliminated; but it forces a modification
of the problem that is being solved.


Use of the GMRES algorithm allows for use of back-projectors $B$ that are not equal
to $A^\top$ without modification of the desired
reconstruction 
model. Furthermore, the algorithm
does not involve any parameters other than the iteration number.
In Sec.\ \ref{sec:ABBA} we present the ABBA framework \cite{hansen2021gmres}
which involves two forms of GMRES called AB-GMRES and BA-GMRES.
In Sec.\ \ref{sec:demo} we demonstrate use of BA-GMRES for unmatched projector/back-projector
pairs and for preconditioning. We conclude this abstract in Sec.\ \ref{sec:conclusion}.

\section{The ABBA framework}
\label{sec:ABBA}

The GMRES algorithm solves a linear system
\begin{equation*}
Sx=v,
\end{equation*}
where
the coefficient matrix
$S$ is a square matrix that is not necessarily symmetric. The relevance for CT image
reconstruction is that
a square non-symmetric matrix arises when multiplying unmatched back-projection $B$
and projection $A$
matrices; i.e., both $AB$ and $BA$ are square non-symmetric matrices.
The original linear system of interest, Eq.\ (\ref{axb}), cannot be directly solved
with GMRES because $A$ is not necessarily square for CT, but this equation can be modified to
\begin{equation}
\label{ABeq}
ABy=b, \quad x = By,
\end{equation}
where the unknown vector $y$ has the same length as the projection data $b$ and the resulting
$m \times m$ 
matrix $AB$ is square.
Also, the normal equations in Eq.\ (\ref{cgnormal}) can be modified by
replacing $A^\top$ with $B$
\begin{equation}
\label{banormal}
B A x = B b,
\end{equation}
and again the
resulting $n \times n$ 
matrix $BA$ is square.
See \cite{elfving2018unmatched} for details.
Modeling CT with Eq.\ (\ref{ABeq}) is similar to the use of
natural pixels \cite{byonocore81,Riddell10,rose2016tv} as the image
is expressed as the back-projection of a sinogram.

We refer to
the GMRES algorithms for solving Eqs. (\ref{ABeq}) and (\ref{banormal})
as AB-GMRES and BA-GMRES, respectively.
The pseudo-code for both algorithms is given in Ref. \cite{hansen2021gmres}, and
we briefly describe the algorithms here. Similar to CGLS,
GMRES is a Krylov subspace method, where the
basis 
vectors of the subspace are generated
by choosing an initial vector and repeatedly applying the
coefficient matrix ($AB$ or $BA$) 
to obtain new linearly independent
basis 
vectors.
For AB-GMRES or BA-GMRES with
a zero initial vector, 
the first
basis vector is
$b$ or $Bb$, respectively, and subsequent
basis 
vectors are generated by applying the matrix
$AB$ or $BA$, respectively.
The GMRES algorithm involves orthonormalization of the Krylov subspace
vectors to obtain a orthonormal basis set that spans the subspace.
At each GMRES iteration the dimension of the subspace is increased by
one and the minimum residual
that can be expressed by the Krylov basis set is found.

The computational burden of GMRES lies with the fact that the
Krylov basis set must be stored
and the number of basis vectors is the same as the number iterations. For AB-GMRES and BA-GMRES
the size of one basis vector is the same as the size of a sinogram and image, respectively.
In our implementation, the basis set is stored on the computer disk. Restart methods \cite{morgan2002gmres}
can reduce the basis vector storage burden, but for this work we demonstrate the basic GMRES
implementation.
\begin{figure}[!t]
\centering
\includegraphics[width=\linewidth]{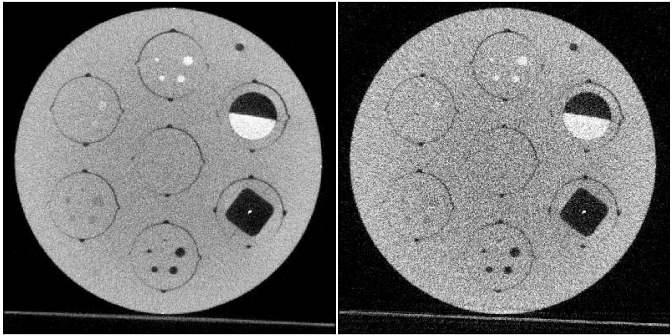}
\caption{Mid-slice images of QA phantom. (Left) FBP reconstructed image
from 720 views. (Right) FBP reconstructed images from 180 views, a 4-fold sub-sampling of
the original CBCT dataset. The grayscale window is [0.,0.25] cm$^{-1}$.}
\label{fig:fbps}
\end{figure}

The AB-GMRES and BA-GMRES algorithms are guaranteed to minimize different data discrepancy measures.
In the case of AB-GMRES, the algorithm minimizes
\begin{equation*}
\|ABy-b\|_2,
\end{equation*}
while BA-GMRES minimizes
\begin{equation*}
\|BAx-Bb\|_2.
\end{equation*}
Note that BA-GMRES is not necessarily minimizing
\begin{equation*}
\|Ax-b\|_2.
\end{equation*}
In this work we focus on BA-GMRES and
we demonstrate
its use 
on cone-beam CT image reconstruction.

\begin{figure}[!t]
\centering
\includegraphics[width=0.9\linewidth]{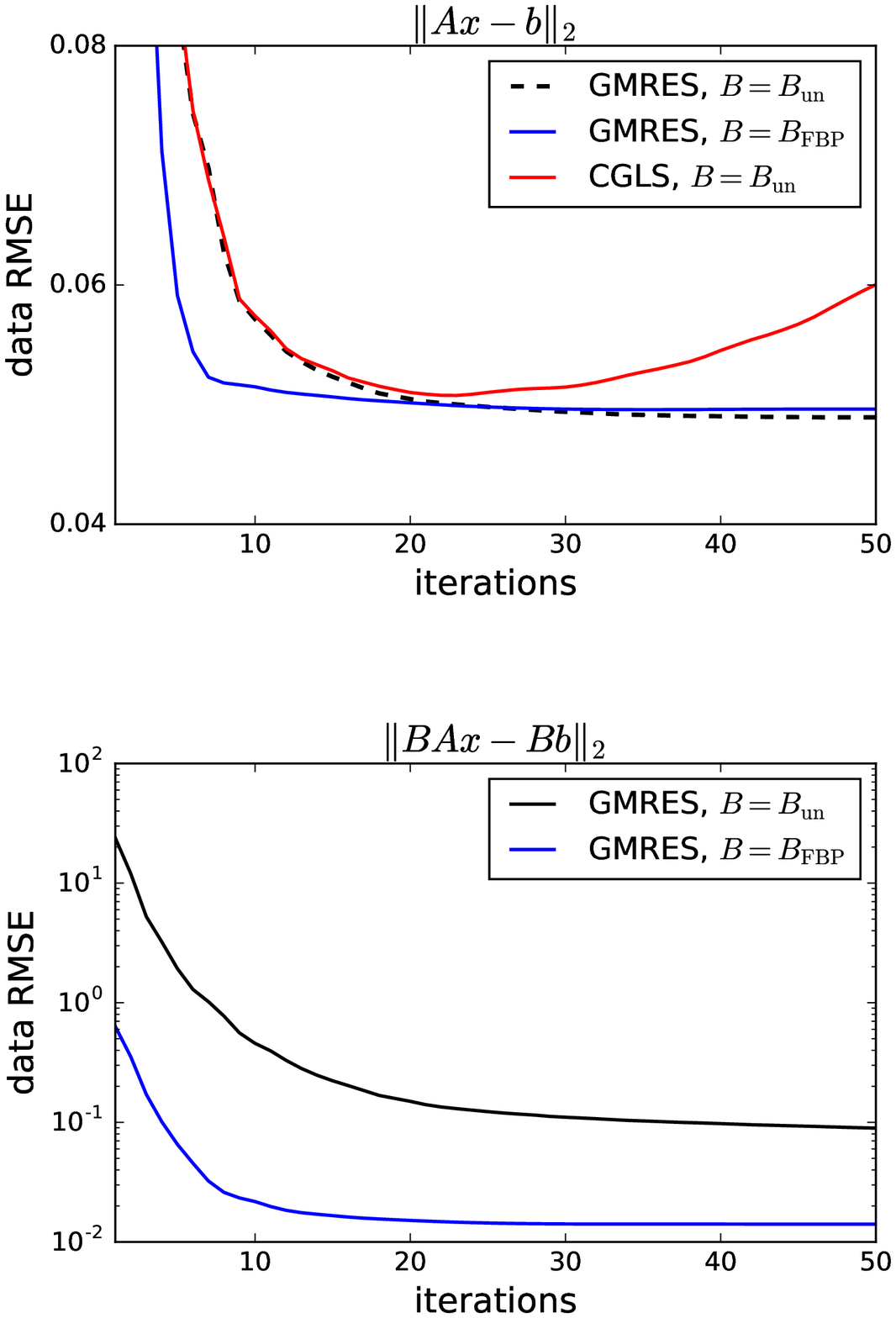}
\caption{Data RMSE in the form of (Top) $\|Ax-b\|_2$ and (Bottom) $\|BAx-Bb\|_2$.
In the top graph unmatched CGLS is also shown to demonstrate divergence of the data RMSE
with unmatched projector/back-projector pairs. The other two curves
correspond to use of voxel-driven back-projection $B=B_{\rm un}$,
and FBP $B=B_{\rm FBP}$.
The projector $A$ is a ray-driven implementation.}
\label{fig:datarmse}
\end{figure}

\section{BA-GMRES applied to cone-beam CT image reconstruction}
\label{sec:demo}

We apply BA-GMRES to a cone-beam CT
(CBCT)
data set acquired on an Epica Pegaso veterinary CT scanner.
The particular scan configuration for the data set is 180 projections taken uniformly over
one circular rotation. The detector size is
$1088 \times 896$ detector pixels, where each pixel
is (0.278mm)$^2$ in size. The 180-view dataset is sub-sampled from a 720-view scan of a quality assurance (QA)
phantom. Image volumes are reconstructed onto a
$1024\times 1024\times 300$ voxel grid, and a reference
volume is generated by use of filtered back-projection (FBP) applied to the full 720-view
dataset and shown in Fig.~\ref{fig:fbps}. Also shown in the figure is FBP applied to the 180-view
sub-sampled dataset.

To demonstrate application of BA-GMRES to CBCT image reconstruction,
we use a ray-driven cone-beam projector
where the matrix elements for $A$ are computed by the line-intersection method.
We consider two implementations
of $B$: (1) $B_{\rm un}$
voxel-driven back-projection using linear interpolation to determine the appropriate projection value
on the detector, and (2) $B_{\rm FBP}=B_{\rm un}F$ filtered back-projection, where $F$ represents
the ramp filter.
See \cite[Chapter~9]{SIAMbook} for details about these discretization models.
The first BA-GMRES implementation tests unmatched
back-projector/projector pairs where $B_\text{un} \approx A^\top$,
and the second implementation includes the additional ramp-filtering
step for preconditioning.
Use of $B_\text{un}$ and $A$ is also shown for CGLS, which requires that $B=A^\top$.

\begin{figure}[!t]
\centering
\includegraphics[width=\linewidth]{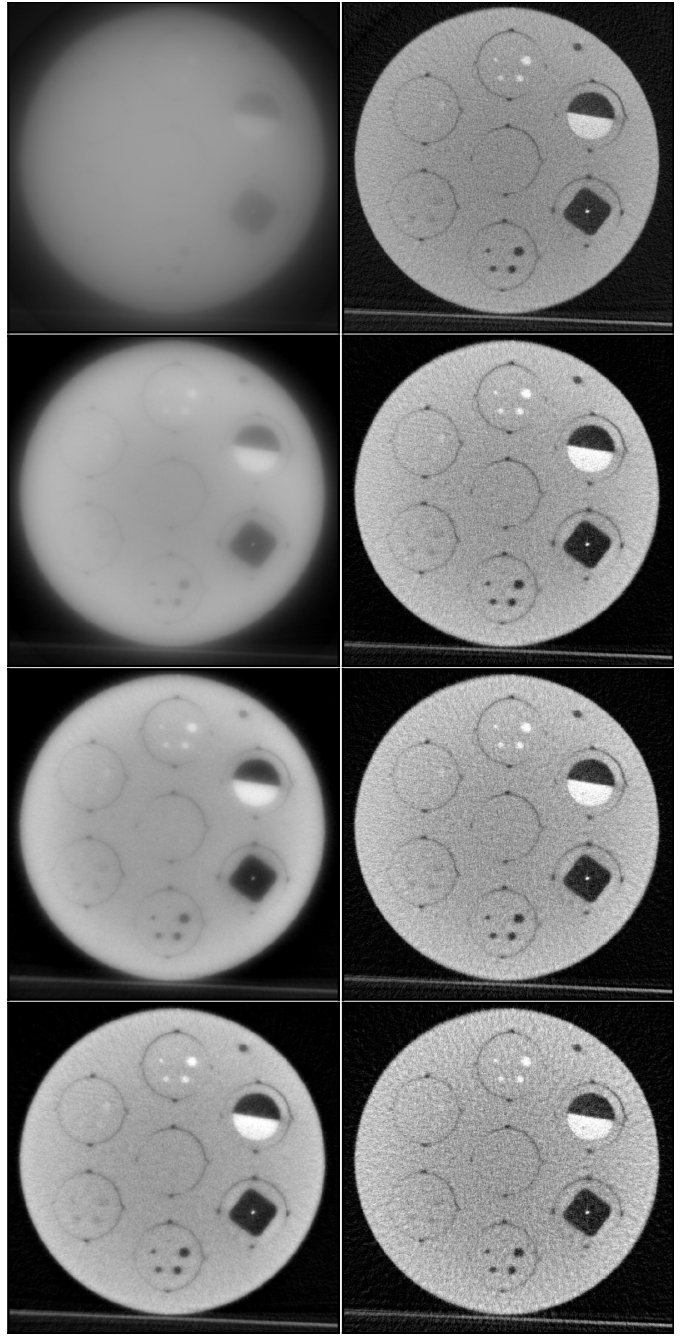}
\caption{Mid-slice BA-GMRES images for 
voxel-driven back-projection $B=B_{\rm un}$ (Left column)
and FBP $B=B_{\rm FBP}$ (Right column).
The shown iteration numbers are 2, 5, 10, and 20 going from the top row to bottom row.
The grayscale window is [0.,0.25] cm$^{-1}$.}
\label{fig:baimages}
\end{figure}

The data root-mean-square-error (RMSE)
curves for both forms of BA-GMRES and CGLS using $B_\text{un}$ and $A$ are shown
in the top panel of Fig.~\ref{fig:datarmse}.
The CGLS result initially shows convergence of the data RMSE, but after 20 iterations the data RMSE
begins to diverge with increasing iteration number, as expected, since
this algorithm is not designed
to work 
for
unmatched 
matrix transpose implementations. The corresponding BA-GMRES
result does show a decreasing data RMSE with iteration number.
For the preconditioned form of BA-GMRES,
the decrease in data RMSE is even more rapid.
The decreasing trends in $\|Ax-b\|_2$ for BA-GMRES occur even though
this algorithm is not guaranteed to reduce this data norm.
Also shown in Fig.~\ref{fig:datarmse} is the data RMSE curves for $\|BAx-Bb\|_2$,
which is guaranteed to decrease with iteration number and they do indeed
show decreasing trends for BA-GMRES.
These issues are elaborated in \cite{hansen2021gmres}.

\begin{figure}[!t]
\centering
\includegraphics[width=0.9\linewidth]{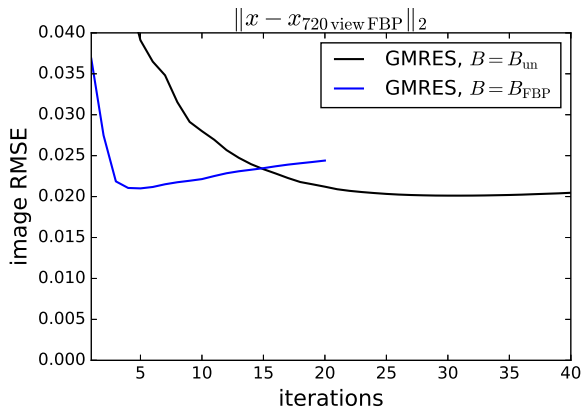}
\caption{Using the 720-view reconstructed volume (see mid-slice image on the left of Fig.~\ref{fig:fbps}
as a reference), the BA-GMRES
reconstructed image RMSE is plotted as a function of iteration number for $B=B_{\rm un}$
and $B=B_{\rm FBP}$.
}
\label{fig:imagermse}
\end{figure}

The mid-slice images for BA-GMRES using both $B$ implementations are shown in Fig.~\ref{fig:baimages} at different
iteration numbers. Preconditioning has a clear effect on the convergence as all the phantom structures are clearly visible
in the early iterations and the gray-level is stabilized already at the fifth iteration. The BA-GMRES result
without preconditioning is also fairly efficient as the main features of the QA phantom are visible at 20 iterations.

\begin{figure}[!t]
\centering
\includegraphics[width=\linewidth]{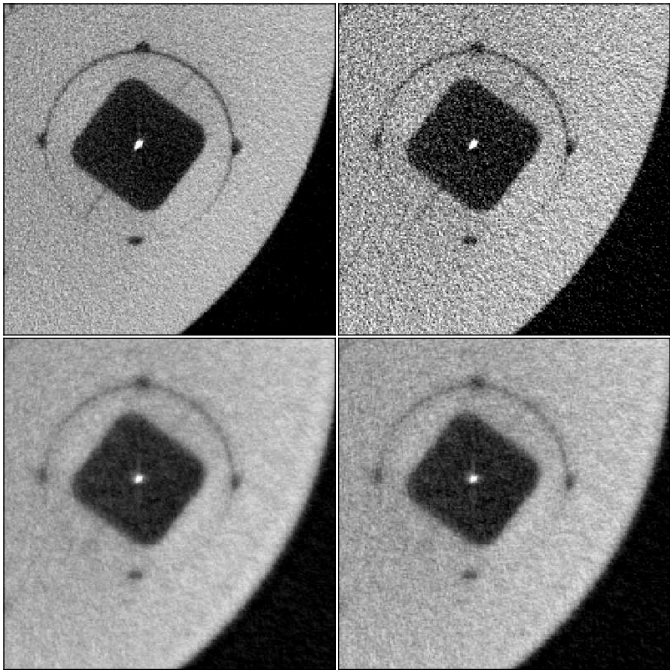}
\caption{Mid-slice ROI images of QA phantom. (Top, Left) FBP reconstructed image
from 720 views. (Top, Right) FBP reconstructed images from 180 views.
(Bottom, Left) BA-GMRES image for $B=B_{\rm un}$ at iteration 29.
(Bottom, Right) BA-GMRES image for $B=B_{\rm FBP}$ at iteration 4.
The grayscale window is [0.,0.25] cm$^{-1}$.}
\label{fig:barois}
\end{figure}

One measure of image quality is to compare the reconstructed volumes to a ground truth image. Employing the 720-view
FBP reconstructed volume as a surrogate for the ground truth, the image RMSE is plotted in Fig.~\ref{fig:imagermse}
for both versions of BA-GMRES. The BA-GMRES implementation with $B=B_\text{un}$ achieves a minimum image
RMSE of 0.0201 at iteration 29, while the preconditioned version with $B=B_\text{un}$
achieves a minimum image RMSE of 0.0210 at iteration 4.
For comparison the 180-view FBP result has an image RMSE of 0.0347. To appreciate the various image qualities, ROI images
of the mid-slice are shown at the minimum image RMSE iteration numbers in Fig.~\ref{fig:barois}.

That the image RMSE has a minimum at finite iteration number is a
well-known phenomenon in
iterative 
image reconstruction and it is known as semi-convergence \cite[Chapter~11]{SIAMbook}.
Early stopping in such algorithms
is a form of regularization because
the components associated with large singular values of $A$
converge fast, while the unwanted noisy components associated with smaller
singular values -- that cause strong image artifacts -- appear after more iterations.
Semi-convergence is observed in the image RMSE curves of Fig.~\ref{fig:imagermse}
and visually in the preconditioned BA-GMRES series of Fig.~\ref{fig:baimages} where the image at 20 iterations
clearly shows strong artifacts from iterating too far. The semi-convergence issue also presents a practical
dilemma for preconditioning. With the shown preconditioned BA-GMRES results, the minimum image RMSE result is
obtained already at the fourth iteration; thus the iteration number provides only coarse control over its image quality.
The un-preconditioned BA-GMRES implementation achieves its image RMSE minimizer at the 29th iteration, which
is computationally less efficient, but on the other hand the iteration number provides a finer control over the image quality.
In any case, the BA-GMRES framework provides a flexible means for implementing back-projectors or preconditioning
schemes, and optimizing the $B$ implementation and iteration number will depend on the imaging task of interest.

\section{Conclusion}
\label{sec:conclusion}

This work presents an iterative image reconstruction framework for linear CT problems that allows for the use
of unmatched back-projector/projector pairs in a straight-forward manner. This possibility is convenient for
implementation of efficient back-projectors, linear modeling of complex physics, and preconditioning. Also, because
it is clear what equation is being solved when $B \neq A^\top$, BA-GMRES can be used for solving linear sub-problems
that may arise in non-linear iterative image reconstruction.  The BA-GMRES algorithm does present a challenge for
computer memory because the Krylov basis set needs to be stored during the iteration, but the present demonstration
on CBCT image reconstruction does show that BA-GMRES can be applied to large-scale CT systems of clinical interest.


%

\section*{Acknowledgment}

The authors wish to thank Holly Stewart and Christopher Kawcak from
Colorado State University for providing the QA phantom data.
This work was supported in part by the Grayson-Jockey Club Research
Foundation, NIH Grant Nos. R01-EB026282 and R01-EB023968,
and a Villum Investigator grant (no. 25893) from The Villum Foundation.
The contents of this article are solely the responsibility of
the authors and do not necessarily represent the official
views of the National Institutes of Health.

\ifCLASSOPTIONcaptionsoff
  \newpage
\fi



%

\end{document}